# Directional Radio Propagation Path Loss Models for Millimeter-Wave Wireless Networks in the 28-, 60-, and 73-GHz Bands

Ahmed Iyanda Sulyman, *Senior Member, IEEE*, Abdulmalik Alwarafy, George R. MacCartney, Jr., *Student Member, IEEE*, Theodore S. Rappaport, *Fellow, IEEE*, and Abdulhameed Alsanie

*Abstract*—Fifth-generation (5G) cellular systems are likely to operate in the centimeter-wave (3–30 GHz) and millimeter-wave (30–300 GHz) frequency bands, where a vast amount of under-utilized bandwidth exists world-wide. To assist in the research and development of these emerging wireless systems, a myriad of measurement studies have been conducted to characterize path loss in urban environments at these frequencies. The standard theoretical free space (FS) and Stanford University Interim (SUI) empirical path loss models were recently modified to fit path loss models obtained from measurements performed at 28 GHz and 38 GHz, using simple correction factors. In this paper, we provide similar correction factors for models at 60 GHz and 73 GHz. By imparting slope correction factors on the FS and SUI path loss models to closely match the close-in (CI) free space reference distance path loss models, millimeter-wave path loss can be accurately estimated (with popular models) for 5G cellular planning at 60 GHz and 73 GHz. Additionally, new millimeter-wave beam combining path loss models are provided at 28 GHz and 73 GHz by considering the simultaneous combination of signals from multiple antenna pointing directions between the transmitter and receiver that result in the strongest received power. Such directional channel models are important for future adaptive array systems at millimeter-wave frequencies.

*Index Terms*—Radio propagation, path loss, 5G cellular, millimeter-wave, close-in free space reference distance, SUI, beam combining.

## I. Introduction

FIFTH-generation (5G) cellular systems are expected to exploit the upper centimeter-wave (cmWave) and mmWave frequency bands (3-300 GHz) that contain abundant amounts of unused spectrum capable of delivering multi-gigabits-per second data rates for mobile and backhaul applications [1]–[3]. Radio propagation at these frequencies, when compared with unity gain antennas, suffers higher free space path loss (FSPL) in the first meter of propagation from a transmit antenna when compared to frequencies below 6 GHz, due to the increase in carrier frequency as governed by Friis' FSPL equation: $\text{FSPL}(f,d)[\text{dB}] = 20\log_{10}((4\pi df)/(3\times 10^8))$ [4], [5]. Thus, the use of high gain antennas at both ends of the link are envisioned to overcome the greater path loss, and will use beamforming and beam combining techniques to enhance link quality with interference rejection [3], [5]–[9]. Candidate carrier frequencies for 5G in the cmWave and mmWave bands are 15 GHz, 28 GHz, 38 GHz, 60 GHz, and 73 GHz, with particular interest in the 28 and 38 GHz bands [10], the unlicensed 60 GHz band, and the 70-80 GHz E-band spectrum region [10]–[14], [16].

A number of measurement campaigns and studies were conducted at frequencies above 6 GHz that generated large-scale path loss models [14], [16], [17]–[20] and spatial/temporal channel statistics [15], [30] useful for mmWave channel modeling, similar to past empirically-based mobile channel models based upon measurements at microwave frequencies [21], [22]. Path loss models can be used in modulation and coding scheme designs, power budget calculations and cellular coverage/interference predictions. The importance of a physical basis in such models is proven in [14], [15], [23], and [25] and is easily implemented by using a 1 meter (m) close-in free space reference distance that characterizes FSPL in the first meter of propagation before the signal encounters any obstructions in line-of-sight (LOS) or non-line-of-sight (NLOS) scenarios [14], [16], [23]. Such models have an advantage in stability and accuracy over vast use cases since they tie path loss at different transmitter (TX) and receiver (RX) separation distances to the close-in, frequency dependent free space path loss, which is accurately determined over the entire frequency range at a close-in 1 m distance using Friis' free space path loss model [14]–[16], [23]–[25]. This close-in reference distance modeling method ensures stability and repeatability across a broad spectrum of measurements, distances, locations, environments, and/or frequencies, with consistent accuracy and stability outside of the measurement range by tying path loss to free space at a close-in distance [23], [25].

Modifications to the well-known FS and the SUI path loss models were implemented in [10] to make them suitable for path loss estimation at 28 GHz and 38 GHz, by computing slope correction factors to match them to empirically-based close-in (CI) free space reference distance path loss models. Slope correction factors have not yet been publicly available

Manuscript received January 6, 2016; revised May 22, 2016; accepted July 12, 2016. Date of publication July 26, 2016; date of current version October 7, 2016. This work was supported in part by the National Plan for Science, Technology and Innovation (MAARIFAH), King Abdulaziz City for Science and Technology, Kingdom of Saudi Arabia, Award Number 14-ELE871-02, in part by the GAANN Fellowship Program, and in part by the Three National Science Foundation (NSF) under Grant 1320472, Grant 1302336, and Grant 1555332. The associate editor coordinating the review of this paper and approving it for publication was J. Wu.

A. I. Sulyman, A. Alwarafy, and A. Alsanie are with the Department of Electrical Engineering, King Saud University, Riyadh 12372, Saudi Arabia (e-mail: asulyman@ksu.edu.sa; 432108488@ksu.edu.sa; sanie@ksu.edu.sa).

G. R. MacCartney, Jr., and T. S. Rappaport are with the New York University Tandon School of Engineering, Brooklyn, NY 11201 USA (e-mail: gmac@nyu.edu; tsr@nyu.edu).

Color versions of one or more of the figures in this paper are available online at http://ieeexplore.ieee.org.

Digital Object Identifier 10.1109/TWC.2016.2594067

for 60 GHz and 73 GHz bands; therefore, in this paper we provide such correction factors that modify the FS and SUI models [14], [17], [19], [20], keeping in mind that 60 GHz will suffer an additional 10-20 dB per km atmospheric loss as compared to other mmWave bands, and that rain attenuation is also a function of frequency up to about 100 GHz [2], [24].

Due to the natural increased attenuation at mmWave frequencies compared to traditional cellular frequencies due to FSPL in the first meter of propagation (when compared using identical gain antennas), beamforming and beam combining methods have been envisioned to mitigate path loss, to increase coverage distance, and consequently increase the signal-to-noise ratio (SNR) at the receiver which in turn will support greater data rates (higher order modulations) [6]–[10], [14], [18], [24], [26], [38], [40]. Beam combining practices are effective in NLOS scenarios where signals can be weak due to obstructions in the environment, but where energy still reaches the receiver by scattering and reflection at many angles of arrival. In this work, new mmWave beam combining path loss models are provided that explicitly account for the number of beams (or discrete angles) combined at the receiver, within the path loss model equation.

The rest of this paper is organized as follows: Section II introduces the modified FS, SUI, and CI beam combining path loss models, Section III describes the measurements and modified FS and SUI path loss models and parameters for 60 GHz and 73 GHz, Section IV provides the parameters for the modified CI directional beam combining path loss models at 28 GHz and 73 GHz, and concluding remarks are given in Section V.

## II. MODIFIED PATH LOSS MODELS

### A. Modified FS and SUI Path Loss Models

Path loss (PL) for 3G and 4G cellular networks operating above 2 GHz in the microwave bands can be estimated using the SUI model for IEEE 802.16e systems [21]. The standard SUI model is given below [10], [21] where $PL_{SUI}$ in dB is found by:

$$PL_{SUI}(d)[dB] = FSPL(f, 1\text{ m})[dB] + 10n \log_{10}\left(\frac{d}{1\text{ m}}\right)$$
$$+ X_{f_C} + X_{RX} + X_\sigma \quad (1)$$

where:

$$FSPL(f, 1\text{ m})[dB] = 20 \cdot \log_{10}\left(\frac{4\pi f}{3 \times 10^8}\right)$$
$$= 32.4 + 20 \cdot \log_{10}(f_{GHz}) \quad (1a)$$

$$n = a - b \cdot h_{TX}(\text{m}) + \frac{c}{h_{TX}(\text{m})} \quad (1b)$$

$$X_{f_c} = 6 \cdot \log_{10}\left(\frac{f_{MHz}}{2000}\right), \ f > 2\text{ GHz} \quad (1c)$$

$$X_{RX} = -10.8 \cdot \log_{10}\left(\frac{h_{RX}(\text{m})}{2}\right),$$
$$\text{for terrain types A and B} \quad (1d)$$

$$X_{RX} = -20 \cdot \log_{10}\left(\frac{h_{RX}(\text{m})}{2}\right),$$
$$\text{for terrain type C} \quad (1e)$$

where $f$ is the carrier frequency in Hertz, $f_{GHz}$ and $f_{MHz}$ denote the carrier frequency when written in GHz and MHz, respectively. FSPL($f$, 1 m) in (1a) denotes the free space path loss in dB at 1 m, $X_{f_c}$ and $X_{RX}$ in (1c), (1d), and (1e) denote the correction factors for frequency and receiver heights, respectively, and $X_\sigma$ in (1) is the typical log-normal random shadowing variable with 0 dB mean and standard deviation $\sigma$ that ranges between 8.2 dB and 10.6 dB [10]. $f_{MHz}$ in (1c) is the carrier frequency in MHz, and $h_{TX}$ and $h_{RX}$ denote the TX and RX antenna heights in meters, respectively. The parameters $a$, $b$, and $c$ in (1b) are constants used to model typical terrain types [10], [21]. For Terrain type A (hilly, dense vegetation), $a = 4.6$, $b = 0.0075$, and $c = 12.6$. For Terrain type B (hilly, rare vegetation), $a = 4.0$, $b = 0.0065$, and $c = 17.1$. For Terrain type C (flat, rare vegetation), $a = 3.6$, $b = 0.005$, and $c = 20$.

For cases where there is a clear line-of-sight (LOS) path between the TX and the RX, Friis' FSPL model [4] is used to estimate path loss as a function of distance while using theoretical free space path loss at 1m as a physical anchor which is valid over all frequencies throughout the cmWave and mmWave bands [25]. The FS equation explicitly uses a path loss exponent (PLE) of 2.0 to compute $PL_{FS}$ in dB and is given by:

$$PL_{FS}(d)[dB] = -10 \cdot \log_{10}\left(\frac{G_t G_r \lambda^2}{(4\pi d)^2}\right) \quad (2)$$

where $\lambda$ is the carrier wavelength, and $G_t$ and $G_r$ denote the TX and RX antenna gains, respectively. In the most general form, the CI path loss model for computing path loss, $PL_{CI}$ is given in Eq. (3):

$$PL_{CI}(f_{GHz}, d)[dB] = 32.4 + 20 \cdot \log_{10}(f_{GHz})$$
$$+ 10n \log_{10}\left(\frac{d}{1\text{ m}}\right) + X_\sigma \quad (3)$$

with theoretical FSPL at 1 m ($32.4 + 20 \cdot \log_{10}(f_{GHz})$) as a reference and the PLE ($n$) that describes the attenuation per decade of log-distance beyond 1 meter [14], [24], [27], [28].

For LOS environments, the $PL_{FS}$ path loss equation given in (2) is modified with the slope correction factor $\alpha_{LOS}$ to best match the LOS $PL_{CI}$ path loss model (based on measured data) as described in [10], where the modified free space model is given by $PL_{FS,\text{mod}}$ in dB:

$$PL_{FS,Mod}(d)[dB] = \alpha_{LOS} \times \left(PL_{FS}(d) - PL_{FS}(d_0)\right)$$
$$+ PL(d_0) + X_\sigma \text{ with } d_0 = 1 \text{ meter} \quad (4)$$

For NLOS environments, the modified SUI model of equation (1) is used with a slope correction factor that results in a model that best fits a close-in (CI) reference distance with $d_0 = 1$ meter using the $PL_{CI}$ NLOS path loss model (based on measured data). The modified SUI model for NLOS is given as [10]:

$$PL_{SUI,Mod}(d)[dB] = \alpha_{NLOS} \times (PL_{SUI}(d) - PL_{SUI}(d_0))$$
$$+ PL(d_0) + X_\sigma \quad (5)$$

where $\alpha_{NLOS}$ is the slope correction factor that modifies the SUI model of Eq. (1) [21] to closely match the NLOS CI

model (3). Simply put, the slope correction factor is applied to the original SUI model so as to provide a modified SUI model that matches the empirically determined 1 m close-in free space reference distance path loss model, $PL_{CI}$, obtained from measured data.

### B. Directional Beam Combining Path Loss Models at mmWave Bands

A new general equation for calculating directional path loss when combining unique directional beams with the strongest received power was proposed in [29], that effectively scales a directional PLE (formed using the single best beam) from the CI model in (3) based on the number of beams considered for combining. To the best of the authors' knowledge, the current paper represents the first time in the literature that the number of combined beams has been incorporated directly into directional path loss equations at mmWave bands.

Assuming that receivers can be built to properly align and coherently combine (CC) and non-coherently combine (NCC) the received energy in wireless transmissions from different antenna beam patterns simultaneously, the following equations determine how the energy (or power) is combined [14], [26], [38], [40]:

$$P_{CC} = \left(\sum_{i=1}^{N_r} \sqrt{P_i}\right)^2 \quad (6)$$

$$P_{NCC} = \sum_{i=1}^{N_r} P_i \quad (7)$$

where $N_r$ is the number of unique orthogonal angles (beams) used for combining received power for a specific TX-RX communication link and $P_i$ is the received power in linear units (e.g., mW) for each unique beam angle. By combining the energy (or power) of the same signal at different angles of arrival it is possible to reduce the path loss exponent (PLE) of the aggregate received signal, and to further reduce the PLE by increasing the number of unique angles for beam combining. Work in [14] provided individual PLEs based on the number of aggregate beams used for combining. However, a simpler and more general mmWave beam combining close-in free space reference distance (BC-CI) path loss model is developed here for NLOS urban environments that accounts for the number of combined angles (beams) at the receiver, while using only three parameters: PLE of the single best (strongest) beam, $A$ the weighting factor, and $N_r$ the number of beams used for combining. In this new propagation model for directional beam combining, the PLE of the single best beam and $A$ remain fixed, while the number of beams $N_r$ is the only input parameter to account for the range of available beam combinations (orthogonal simultaneous receiving directions) in the channel.

The reduction in the aggregate PLE when combining beams at 28 GHz and 73 GHz shows that the path loss exponent for each TX-RX combination decreases logarithmically as a function of the number of combined beams, for both the CC and NCC schemes [14]. According to this trend, we propose a new beam combining path loss model that uses a 1 m close-in FSPL reference, i.e., the BC-CI path loss model (modified from (3) to include the number of simultaneously combined beams) that incorporates a varying number of available beams as given by:

$$PL_{BC\text{-}CI} = FSPL(f, 1\text{ m})[dB] + \left(10n_{(1\text{ beam})} \log_{10}\left(\frac{d}{1\text{ m}}\right)\right) \times \left(1 - A\log_2(N_r)\right) + X_\sigma \quad (8)$$

where $n_{(1\text{ beam})}$ is the PLE for the single (strongest, or best) beam, $N_r$ is the number of unique pointing beams combined, and $A$ is the weighting factor that is obtained via minimum mean square error (MMSE) for fitting the BC-CI path loss model to the original CI beam combining data [26] with the minimum root mean squared error (RMSE), over each number of beams used for combining. The $(1 - A\log_2(N_r))$ term effectively scales the single beam PLE ($n_{(1\text{ beam})}$) to obtain the respective PLE for each number of beams combined. Note that the model in (8) reduces to (3) with the single best (strongest) beam PLE when $N_r = 1$. Comparisons of this model and the original model proposed in [14] are provided in Section IV, and are remarkably accurate. A specific advantage of the proposed model here is that it is possible to generalize the number of combined beams to $N_r > 4$ when estimating path loss for beam combining beyond the four best beams, although measurements show that with antenna beamwidths greater than 7°, there are seldom more than 5 distinct angular directions of arrival (e.g. seldom more than 5 beams) [14], [15]. As future narrow beam mmWave or THz systems evolve in the coming decades, this model will be able to facilitate system design and performance evaluations for cases where receivers combine all significant energy received from various angles of arrival, in which case $N_r$ can be an arbitrary number [14], [30], [31].

The BC-CI model was developed from measured data in [14] [26] [38] as follows: for a given $n_{(1\text{ beam})}$, $N_r$, and $PL(d_i), i = 1, ..., K$ from the measurement data of $K$ measurement locations, compute the squared-error $J(A) = \sum_{i=1}^{K}[PL(d_i) - PL(d_i)_{BC\text{-}CI}]^2$. The MMSE solution for the $A$ weighting factor is computed by solving $\partial J(A)/\partial A = 0$ for $A$, while the standard deviation $\sigma$ is computed as $\sigma = \sqrt{E\{[PL(d_i) - PL(d_i)_{BC\text{-}CI}]^2\}}$, with $E\{\cdot\}$ denoting the expectation operation.

### III. RESULTS AND PARAMETERS FOR MODIFIED PATH LOSS MODELS AT 60 GHz AND 73 GHz

The 28 GHz and 73 GHz measurements were performed in New York City [2] [10] [14] [19] [32] while the 60 GHz measurements were performed in Austin, Texas [14] [16] [17] [20], all using a broadband sliding correlator channel sounder with a pseudorandom noise (PN) maximal length sequence of length 2047, clocked at a rate of 400 megachips-per-second (Mcps) for the 28 and 73 GHz measurements, and 750 Mcps for the 60 GHz measurements. TX antenna heights of 7 and 17 meters were used at 28 and 73 GHz, with 1.5 m (mobile) RX heights for 28 GHz measurements, and 2 m (mobile) and 4.06 m (backhaul) RX heights for 73 GHz measurements.

New York City measurement campaigns consisted of directional high gain horn antennas with 24.5 dBi of gain and 10.9°

TABLE I
SLOPE CORRECTION FACTORS FOR THE MODIFIED FS AND SUI PATH LOSS MODELS AT 60 GHz FOR THE LOS AND NLOS SCENARIOS IN COURTYARD AND IN-VEHICLE ENVIRONMENTS. MODIFIED SUI MODELS ARE GIVEN FOR THREE POTENTIAL TERRAIN TYPES (A, B, AND C) FOR 60 GHz NLOS BASED ON MEASUREMENT DISTANCES BETWEEN 29 m AND 129 m

| | Frequency: 60 GHz | | | |
|---|---|---|---|---|
| Environment | NLOS | | LOS | |
| Environment Details | Courtyard | In-vehicle | Courtyard | In-vehicle |
| TX height (m) | 1.5 | | | |
| RX height (m) | 1.5 | | | |
| $d_0$ (m) | 1 | | | |
| PLE $n$ | 3.6 | 5.4 | 2.2 | 2.5 |
| $\sigma$ (dB) | 9.0 | 14.8 | 2.0 | 3.5 |
| TX Gain (dBi) | 25 | | | |
| TX HPBW (°) | 7.3 | | | |
| RX Gain (dBi) | 25 | | | |
| RX HPBW (°) | 7.3 | | | |
| Slope Correction Factor: $\alpha$ (Terrain A) | 0.277 | 0.416 | | |
| Slope Correction Factor: $\alpha$ (Terrain B) | 0.234 | 0.351 | 1.10 | 1.25 |
| Slope Correction Factor: $\alpha$ (Terrain C) | 0.213 | 0.319 | | |

azimuth half-power beamwidth (HPBW) at 28 GHz, and 27 dBi of gain and 7° azimuth HPBW at 73 GHz. TX and RX antenna sweeps were conducted for various fixed elevation angles in both campaigns for TX-RX links ranging between 31 m to 102 m in LOS scenarios and between 53 m to 187 m in NLOS scenarios in the dense urban environment of New York City [14]. Maps of the measurement environments as well as actual measured data can be found in [32].

The 60 GHz measurements were performed for LOS and NLOS scenarios in a courtyard (peer-to-peer) and from a street into a vehicle (in-vehicle) environments in Austin, Texas, at the University of Texas at Austin campus [14] [16] [17] [20]. LOS scenarios were when the TX and RX antennas were aligned on boresight in the azimuth and elevation planes with a clear optical path between the two. NLOS scenarios were when the TX and RX antennas were separated by obstructions, or when the TX and RX antennas were not aligned on boresight, even if an optical path between the two existed. High gain directional horn antennas with 25 dBi of gain and 7.3° azimuth HPBW TX and RX antennas were used for the measurements with TX and RX heights at 1.5 m (mobile/peer-to-peer) [14]. Various antenna pointing configurations were considered at distances ranging from 29 m to 129 m, with focus on device-to-device (D2D) and large reflectors for antenna orientations. Additional details can be found in [17].

### A. 60 GHz Peer-to-Peer Courtyard and In-Vehicle

The 60 GHz peer-to-peer courtyard and in-vehicle measurement environments are best classified as flat and dense vegetation which is closely modeled by terrain type A of the SUI model with parameters $a = 4.6$, $b = 0.0075$, and $c = 12.6$. The measured path loss data, CI path loss models, and modified FS and modified SUI path loss models for 60 GHz LOS and NLOS scenarios for courtyard and in-vehicle environments are shown in Figs. 1 and 2, with model parameters given in Table I. The standard FS path loss model (for LOS) and microwave SUI model (for NLOS) in Figs. 1 and 2 without the use of correction factors clearly do not match the respective 1 m CI models, nor do they accurately estimate the measured path loss data in general.

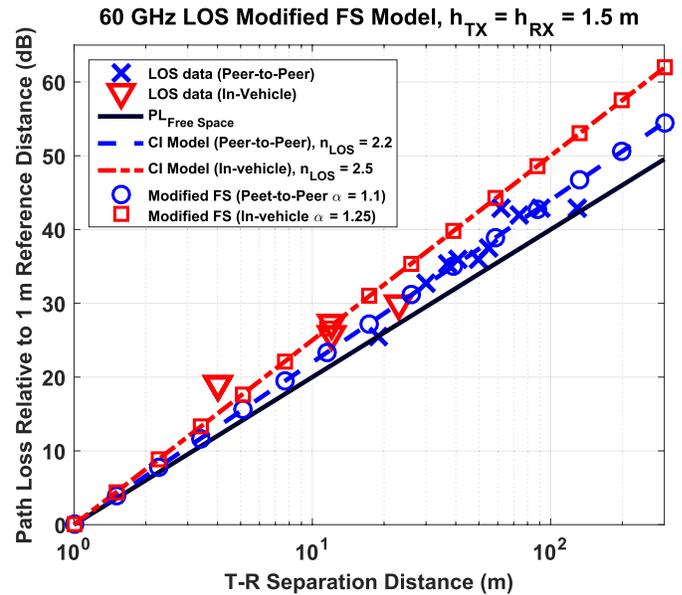

Fig. 1. 1 m CI path loss models with the modified FS path loss model at 60 GHz for LOS peer-to-peer courtyard and in-vehicle environments with $\alpha = 1.1$ for courtyard and $\alpha = 1.25$ for in-vehicle.

The slope correction factors computed via the MMSE approach for LOS ($\alpha_{\text{LOS}}$) and NLOS ($\alpha_{\text{NLOS}}$) scenarios are provided in Table I for the modified FS and SUI models in (4) and (5), and the resulting models better match the trend given by the empirical CI path loss models, as depicted in Figs. 1 and 2, although the fit to NLOS data is not as good as one would want, indicating a path loss model that accounts for car body loss might be more appropriate (see [39, p. 1670]). Table I provides the slope correction factors for the terrain type A for both LOS and NLOS, but also for terrain types B and C in NLOS, since the courtyard and in-vehicle environments for NLOS scenarios may be considered for any terrain type.

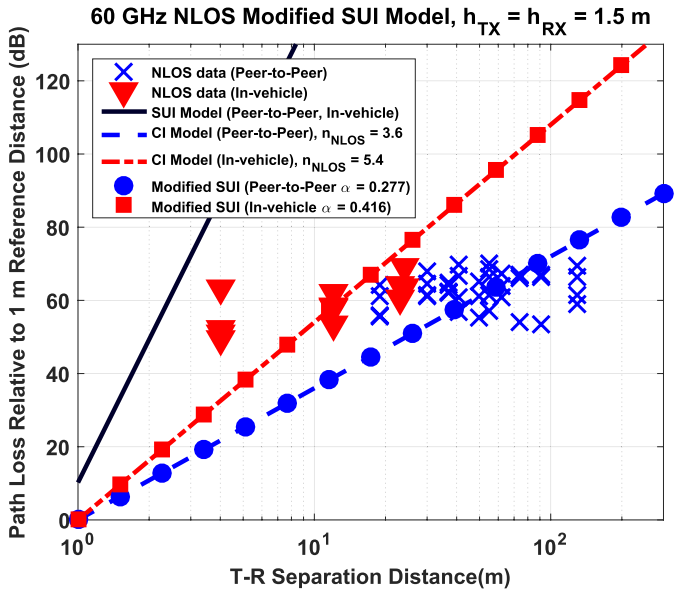

Fig. 2. 1 m CI path loss models along with the modified SUI (terrain A) path loss model at 60 GHz for NLOS peer-to-peer courtyard and in-vehicle environments with $\alpha = 0.277$ for courtyard and $\alpha = 0.416$ for in-vehicle.

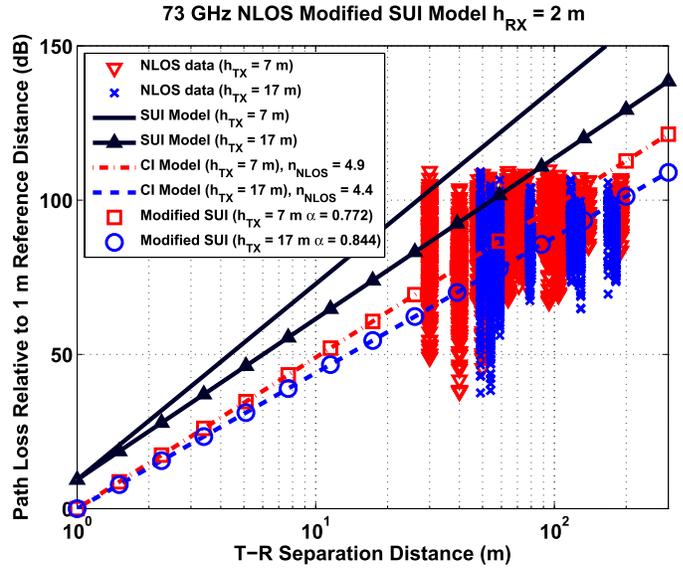

Fig. 4. 1 m CI path loss models along with the modified SUI (terrain A) path loss model for the 73 GHz NLOS scenario with RX heights of 2 m with $\alpha = 0.772$ for $h_{tx} = 7$ m and $\alpha = 0.844$ for $h_{tx} = 17$ m.

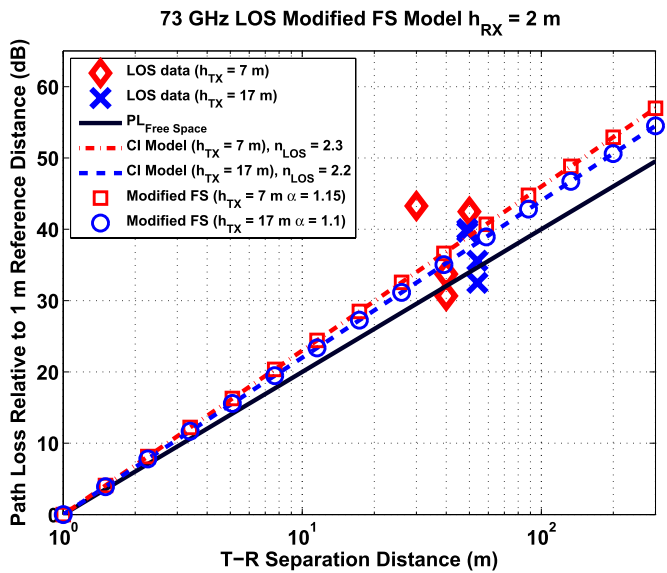

Fig. 3. 1 m CI path loss models along with the modified FS path loss model for the 73 GHz LOS scenario with RX heights of 2 m with $\alpha = 1.15$ for $h_{tx} = 7$ m and $\alpha = 1.1$ for $h_{tx} = 17$ m.

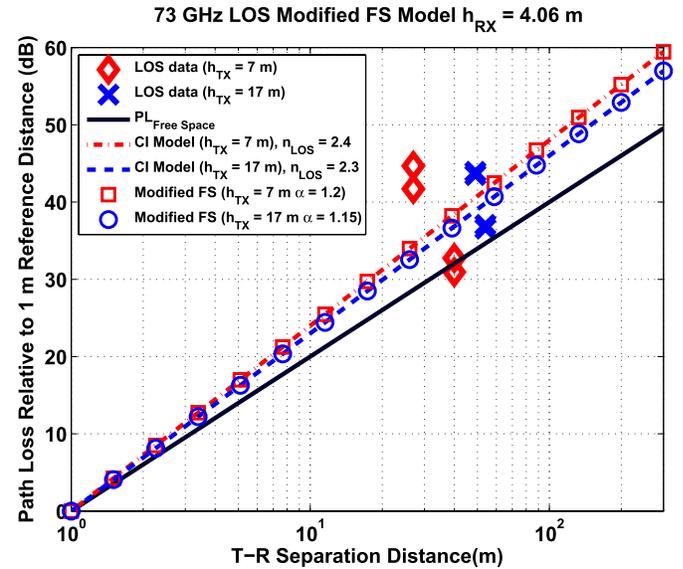

Fig. 5. 1 m CI path loss models along with the modified FS path loss models for the 73 GHz LOS scenario with RX heights of 4.06 m with $\alpha = 1.2$ for $h_{tx} = 7$ m and $\alpha = 1.15$ for $h_{tx} = 17$ m.

### B. 73 GHz Mobile and Backhaul

Modified FS and SUI path loss models are now provided for 73 GHz mobile and backhaul directional measurements for LOS and NLOS scenarios, respectively. Directional CI path loss models are shown in Figs. 3, 4, 5, and 6, with model parameters provided in Table II for differing TX and RX antenna heights. It is apparent in Figs. 3, 4, 5, and 6 that the FS model in LOS and SUI models in NLOS do not accurately estimate the measured 73 GHz path loss nor do they match the empirical CI models. With the use of slope correction factors (provided in Table II), the modified FS and modified SUI models identically match the estimated CI path loss models with respect to a 1 m free space reference distance which accounts for the physics of unobstructed propagation in the first meter from the TX.

Figs. 5 and 6 display the measured directional path loss data at 73 GHz in LOS and NLOS environments for RX heights of 4.06 m (backhaul scenario). The CI model parameters are provided for TX heights of 7 m and 17 m for antenna height diversity. The figures show the inaccuracy of the original FS and SUI model compared to the measured data and empirical CI models at 73 GHz. The slope correction factors (in Table II) effectively modify the FS and SUI models to accurately predict path loss and are well matched to the more stable and physically-motivated 1 m CI path loss models.

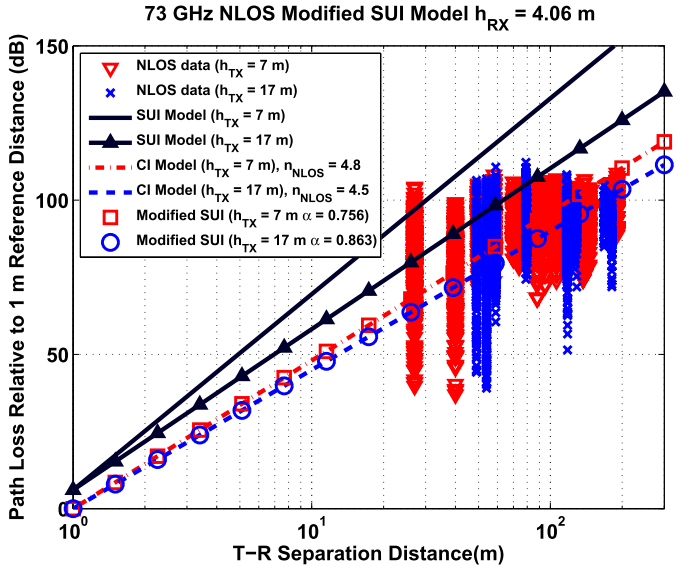

Fig. 6. 1 m CI path loss models along with the modified SUI (terrain A) path loss models for the 73 GHz NLOS scenario with RX heights of 4.06 m with $\alpha = 0.756$ for $h_{tx} = 7$ m and $\alpha = 0.863$ for $h_{tx} = 17$ m.

TABLE II
SLOPE CORRECTION FACTORS FOR THE MODIFIED FS AND SUI PATH LOSS MODELS AT 73 GHz FOR LOS AND NLOS SCENARIOS, FOR TX HEIGHTS OF 7 m AND 17 m, RX HEIGHTS OF 2 m AND 4.06 m, AND FOR TERRAIN TYPE A IN LOS AND TERRAIN TYPES A AND B IN NLOS, BASED ON MEASUREMENT DISTANCES BETWEEN 31 m TO 102 m IN LOS SCENARIOS AND BETWEEN 53 m TO 187 m IN NLOS SCENARIOS IN DOWNTOWN NEW YORK CITY [14]

| Frequency: 73 GHz | | | | | | | | |
|---|---|---|---|---|---|---|---|---|
| Environment | NLOS | | | | LOS | | | |
| TX height (m) | 17 | 7 | 17 | 7 | 17 | 7 | 17 | 7 |
| RX height (m) | 2 | | 4.06 | | 2 | | 4.06 | |
| $d_0$ (m) | 1 | | | | | | | |
| PLE $n$ | 4.4 | 4.9 | 4.5 | 4.8 | 2.2 | 2.3 | 2.3 | 2.4 |
| $\sigma$ (dB) | 11.7 | 11.9 | 12.6 | 12.4 | 4.1 | 6.9 | 4.6 | 9.1 |
| TX Gain (dBi) | 27 | | | | | | | |
| TX HPBW (°) | 7 | | | | | | | |
| RX Gain (dBi) | 27 | | | | | | | |
| RX HPBW (°) | 7 | | | | | | | |
| Slope Correction Factor: $\alpha$ (Terrain type A) | 0.844 | 0.772 | 0.863 | 0.756 | | | | |
| Slope Correction Factor: $\alpha$ (Terrain type B) | 0.899 | 0.766 | 0.919 | 0.75 | 1.1 | 1.15 | 1.15 | 1.2 |

## IV. RESULTS AND PARAMETERS FOR BEAM COMBINING MODEL AT 28 GHz AND 73 GHz

Path loss data for the coherent and non-coherent beam combining of up to the 4 best beams from the 28 GHz and 73 GHz directional measurements in New York City [14] [26] [38] [40] were used to optimize the $A$ weighting factors in (8) via MMSE. The results and optimized parameters are provided in Table III for both 28 GHz and 73 GHz and for coherent and non-coherent power combining methods.

Table III also provides a comparison between the CI PLEs and the BC-CI PLEs computed using the best-fit $A$ weighting factors and the number of antennas $N_r$. The new BC-CI model is consistent with the CI model, with nearly identical PLEs and practically zero RMSE between comparative entries, while both models are grounded in the fundamental physics of radio wave propagation with a frequency-dependent close-in 1 m free space reference distance.

From Table III it is apparent that the $A$ weighting factors are constant for any number of beams combined, given a particular measurement setup (TX/RX heights, carrier frequency, and combining type). This property of the proposed model is very attractive, as it makes it possible to estimate path loss for the general case when the number of combined beams, $N_r$, is more than four. For example, in the case where $N_r = 2, 4, 6, 8$ and 10, etc., one would use $A = 0.0671$ from Table III together with the PLE of the strongest beam, $n_{(1 \text{ beam})}$, along with the respective $N_r$ to estimate path loss for the coherent combining (CC) case, for the corresponding TX/RX heights and the 28 GHz frequency in Table III. Fig. 7 illustrates estimated path loss for cases with $N_r = 1, 2, 4, 6, 8$, and 10. Notice that the CI models obtained from measured data in [14] are also included for the cases of $N_r = 2$ and 4, and they match perfectly with the respective BC-CI models. Only the $A$ weighting values need modification for other specifications (such as carrier frequency or TX/RX heights) when using the BC-CI model, whereas the CI model requires a slightly different PLE for each number of beams considered for combining. It is apparent from Fig. 7 that beam combining provides reduction in path loss for $N_r = 2$ and $N_r = 4$, with diminishing returns when $N_r > 4$. For example for cases with $N_r = 2, 4, 6, 8$ and 10 in Fig. 7, the respective PLEs are 3.56, 3.30, 3.15, 3.05, and 2.96 compared to a PLE of 3.81 for the case of $N_r = 1$ (single best beam only). This result is in agreement with what is established in the literature on SNR/capacity enhancements using diversity combining schemes [33]–[37].

It is also observed from Table III that both the mobile ($h_{RX} = 2$ m) and backhaul ($h_{RX} = 4.06$ m) scenarios at 73 GHz result in almost identical path loss estimates between the original CI models and the new BC-CI models, where the effective PLEs are within two decimal places of each other, which would result in a difference of only approximately 0.1 dB of attenuation per decade of distance between the models (not discernible in practical field measurements). Using the $A$ weighting values in the proposed BC-CI model, which is fixed for any number of beams combined for a particular measurement setup, it is now possible to accurately estimate beam combining path loss for any arbitrary number of beams. Such a model is especially useful when receivers aggregate all available energy received from various angles of arrival in mmWave channels, and may have importance for future massive MIMO or massive multibeam systems at higher frequencies.

From Table III, it is also apparent and not surprising that coherent combination of beams results in a more pronounced reduction in the PLE than non-coherent combination of beams, even though both methods show considerable improvements in increasing the received power. At 28 GHz, the PLE is reduced by approximately 0.5 and 0.3 by coherently and non-coherently adding four beams, respectively, meaning that path loss estimates reduce by 5 dB and 3 dB per decade of log-distance through beam combining. At 73 GHz, coherently combining the 4 best beams compared to the single best beam reduces the PLE from 3.728 and 3.823 to 3.226 and 3.348 for the mobile and backhaul scenarios, respectively. Therefore, the path loss observed at 100 m when considering the single best beam would be achieved at over double the distance of approximately 205 m for the mobile scenario when combining the 4 best beams [38]. This is a reduction in attenuation of approximately 5 dB per decade of distance for both scenarios, a significant improvement that will be useful for E-band systems.

TABLE III
DIRECTIONAL CI AND BC-CI PATH LOSS MODEL PARAMETERS AT 28 GHz AND 73 GHz FOR NLOS SCENARIOS IN NEW YORK CITY WHEN CONSIDERING THE BEST POINTING ANGLES BETWEEN THE TX AND RX. TX HEIGHTS WERE 7 m AND 17 m AT 28 GHz AND 73 GHz, AND RX HEIGHTS WERE 1.5 m AT 28 GHz, AND 2 m (MOBILE) AND 4.06 m (BACKHAUL) AT 73 GHz. TX AND RX ANTENNAS HAD 10.9° AZIMUTH HPBW WITH 24.5 dBi OF GAIN AT 28 GHz, WHILE THE TX AND RX ANTENNAS HAD 7° AZIMUTH HPBW WITH 27 dBi OF GAIN AT 73 GHz

| 28 GHz Directional Beam Combining ($h_{RX}$ = 1.5 m) | | | | | | | | |
|---|---|---|---|---|---|---|---|---|
| | Coherent (CC) | | | | Non-Coherent (NCC) | | | |
| # of Beams ($N_r$) | 1 | 2 | 3 | 4 | 1 | 2 | 3 | 4 |
| $A$ | 0.0671 | | | | 0.0297 | | | |
| $n_{(1\ \text{beam})} \cdot \left(1 - A \log_2(N_r)\right)$ | 3.812 | 3.557 | 3.407 | 3.301 | 3.812 | 3.699 | 3.633 | 3.586 |
| PLE from [14] | 3.812 | 3.548 | 3.406 | 3.307 | 3.812 | 3.692 | 3.631 | 3.591 |
| $\sigma_{BC-CI}$ [dB] | 9.1 | 9.1 | 9.2 | 9.2 | 9.1 | 9.2 | 9.2 | 9.2 |
| $\sigma$ [dB] from [14] | 9.1 | 9.1 | 9.2 | 9.2 | 9.1 | 9.2 | 9.2 | 9.2 |
| $\Delta_\sigma = \|\sigma - \sigma_{BC-CI}\|$ [dB] | 0 | 0 | 0 | 0 | 0 | 0 | 0 | 0 |
| 73 GHz Directional Beam Combining for Mobile ($h_{RX}$ = 2 m) | | | | | | | | |
| | Coherent (CC) | | | | Non-Coherent (NCC) | | | |
| # of Beams ($N_r$) | 1 | 2 | 3 | 4 | 1 | 2 | 3 | 4 |
| $A$ | 0.0673 | | | | 0.0284 | | | |
| $n_{(1\ \text{beam})} \cdot \left(1 - A \log_2(N_r)\right)$ | 3.728 | 3.477 | 3.330 | 3.226 | 3.728 | 3.622 | 3.560 | 3.516 |
| PLE from [14] | 3.728 | 3.466 | 3.327 | 3.235 | 3.728 | 3.613 | 3.557 | 3.523 |
| $\sigma_{BC-CI}$ [dB] | 7.6 | 7.3 | 7.2 | 7.2 | 7.6 | 7.4 | 7.3 | 7.3 |
| $\sigma$ [dB] from [14] | 7.6 | 7.3 | 7.2 | 7.2 | 7.6 | 7.4 | 7.3 | 7.3 |
| $\Delta_\sigma = \|\sigma - \sigma_{BC-CI}\|$ [dB] | 0 | 0 | 0 | 0 | 0 | 0 | 0 | 0 |
| 73 GHz Directional Beam Combining for Backhaul ($h_{RX}$ = 4.06 m) | | | | | | | | |
| | Coherent (CC) | | | | Non-Coherent (NCC) | | | |
| # of Beams ($N_r$) | 1 | 2 | 3 | 4 | 1 | 2 | 3 | 4 |
| $A$ | 0.0621 | | | | 0.0256 | | | |
| $n_{(1\ \text{beam})} \cdot \left(1 - A \log_2(N_r)\right)$ | 3.823 | 3.586 | 3.447 | 3.348 | 3.823 | 3.726 | 3.668 | 3.628 |
| PLE from [14] | 3.823 | 3.578 | 3.446 | 3.353 | 3.823 | 3.718 | 3.667 | 3.632 |
| $\sigma_{BC-CI}$ [dB] | 8.9 | 8.5 | 8.1 | 7.8 | 8.9 | 8.6 | 8.3 | 8.1 |
| $\sigma$ [dB] from [14] | 8.9 | 8.5 | 8.1 | 7.8 | 8.9 | 8.6 | 8.3 | 8.1 |
| $\Delta_\sigma = \|\sigma - \sigma_{BC-CI}\|$ [dB] | 0 | 0 | 0 | 0 | 0 | 0 | 0 | 0 |

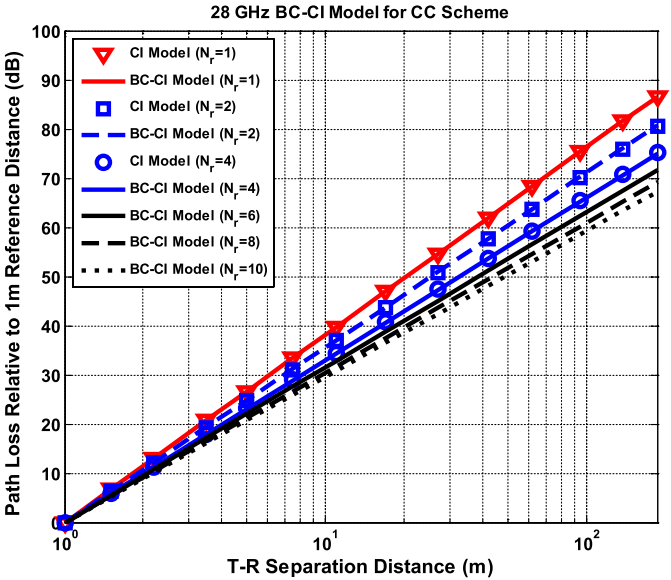

Fig. 7. Path loss models for arbitrary number of beams combined using the proposed BC-CI model at 28 GHz for NLOS scenarios with TX heights of $h_{tx} = 7$ m and $h_{tx} = 17$ m and RX heights of $h_{rx} = 1.5$ m, compared alongside the traditional CI model.

V. CONCLUSION

This paper provides correction factors for modifying the FS and SUI path loss models to fit CI path loss models using a 1 m free space reference distance as obtained from empirical data at 60 GHz and 73 GHz, for both LOS and NLOS scenarios. The modified propagation path loss models are suitable for 5G cellular planning at the 60 GHz and 73 GHz mmWave bands and for estimating mmWave path loss as a function of distance and other environmental and system specific parameters.

Also, a new beam combining close-in (BC-CI) path loss model was proposed based on 28 GHz and 73 GHz directional measurements. The new beam combining model uses three parameters: the single best beam PLE, a weighting factor $A$, and the number of beams as input parameters to estimate distance dependent path loss. The new BC-CI models yield virtually identical RMSE standard deviation as compared to original CI models, therefore validating the accuracy of the new model. Thus, one may extend the number of beams used in the proposed model beyond the number used in this paper (four). This is a particular advantage of the proposed model, as it provides a generalized beam combining model for mmWave path loss prediction. The modified and new path loss models in this paper are all physically-anchored to a 1 m free space reference distance (tied to the true transmit power), where for either LOS or NLOS scenarios the TX will more than likely be free of obstructions in the first meter of propagation. The advantage of using the close-in free space reference distance path loss model with respect to a 1 m reference distance ($d_0$) is the proven stabilty and simplicity, which is further exploited in the resulting modified models presented here. The use of a standardized close-in reference distance of 1 m allows researchers to directly compare their models to drive the standardization of 5G mmWave channel models.

REFERENCES

[1] Z. Pi and F. Khan, "An introduction to millimeter-wave mobile broadband systems," *IEEE Commun. Mag.*, vol. 49, no. 6, pp. 101–107, Jun. 2011.

[2] T. S. Rappaport et al., "Millimeter wave mobile communications for 5G cellular: It will work!" *IEEE Access*, vol. 1, pp. 335–349, May 2013.

[3] S. Rangan, T. S. Rappaport, and E. Erkip, "Millimeter-wave cellular wireless networks: Potentials and challenges," *Proc. IEEE*, vol. 102, no. 3, pp. 366–385, Mar. 2014.

[4] T. S. Rappaport, *Wireless Communications: Principles and Practice*, 2nd ed. Upper Saddle River, NJ, USA: Prentice-Hall, 2002.

[5] H. Shokri-Ghadikolaei, C. Fischione, G. Fodor, P. Popovski, and M. Zorzi, "Millimeter wave cellular networks: A MAC layer perspective," *IEEE Trans. Commun.*, vol. 63, no. 10, pp. 3437–3458, Oct. 2015.

[6] S. Sun, T. S. Rappaport, R. W. Heath, Jr., A. Nix, and S. Rangan, "MIMO for millimeter-wave wireless communications: Beamforming, spatial multiplexing, or both?" *IEEE Commun. Mag.*, vol. 52, no. 12, pp. 110–121, Dec. 2014.

[7] T. Bai, A. Alkhateeb, and R. W. Heath, Jr., "Coverage and capacity of millimeter-wave cellular networks," *IEEE Commun. Mag.*, vol. 52, no. 9, pp. 70–77, Sep. 2014.

[8] A. Alkhateeb, J. Mo, N. González-Prelcic, and R. W. Heath, Jr., "MIMO precoding and combining solutions for millimeter-wave systems," *IEEE Commun. Mag.*, vol. 52, no. 12, pp. 122–131, Dec. 2014.

[9] T. Bai and R. W. Heath, Jr., "Coverage and rate analysis for millimeter-wave cellular networks," *IEEE Trans. Wireless Commun.*, vol. 14, no. 2, pp. 1100–1114, Feb. 2015.

[10] A. I. Sulyman, A. T. Nassar, M. K. Samimi, G. R. MacCartney, Jr., T. S. Rappaport, and A. Alsanie, "Radio propagation path loss models for 5G cellular networks in the 28 GHz and 38 GHz millimeter-wave bands," *IEEE Commun. Mag.*, vol. 52, no. 9, pp. 78–86, Sep. 2014.

[11] C. Dehos, J. L. González, A. De Domenico, D. Kténas, and L. Dussopt, "Millimeter-wave access and backhauling: The solution to the exponential data traffic increase in 5G mobile communications systems?" *IEEE Commun. Mag.*, vol. 52, no. 9, pp. 88–95, Sep. 2014.

[12] *FCC Report and Order and Further Notice of Rule Making, FCC 16-89*, accessed on Jul. 14, 2016. [Online]. Available: http://transition.fcc.gov/Daily_Releases/Daily_Business/2016/db0728/FCC-16-89A1.pdf

[13] Federal Communications Commission. (Jan. 2015). *FCC-14-177*. [Online]. Available: http://apps.fcc.gov/ecfs/proceeding/view?name=14-177

[14] T. S. Rappaport, G. R. MacCartney, Jr., M. K. Samimi, and S. Sun, "Wideband millimeter-wave propagation measurements and channel models for future wireless communication system design (invited paper)," *IEEE Trans. Commun.*, vol. 63, no. 9, pp. 3029–3056, Sep. 2015.

[15] M. K. Samimi and T. S. Rappaport, "3-D millimeter-wave statistical channel model for 5G wireless system design," *IEEE Trans. Microw. Theory Techn.*, vol. 64, no. 7, pp. 2207–2225, Jul. 2016.

[16] G. R. Maccartney, Jr., T. S. Rappaport, S. Sun, and S. Deng, "Indoor office wideband millimeter-wave propagation measurements and channel models at 28 and 73 GHz for ultra-dense 5G wireless networks," *IEEE Access*, vol. 3, pp. 2388–2424, Dec. 2015.

[17] E. Ben-Dor, T. S. Rappaport, Y. Qiao, and S. J. Lauffenburger, "Millimeter-wave 60 GHz outdoor and vehicle AOA propagation measurements using a broadband channel sounder," in *Proc. IEEE Global Telecommun. Conf. (GLOBECOM)*, Dec. 2011, pp. 1–6.

[18] M. R. Akdeniz et al., "Millimeter wave channel modeling and cellular capacity evaluation," *IEEE J. Sel. Areas Commun.*, vol. 32, no. 6, pp. 1164–1179, Jun. 2014. [Online]. Available: http://dx.doi.org/10.1109/JSAC.2014.2328154

[19] G. R. MacCartney, Jr., and T. S. Rappaport, "73 GHz millimeter wave propagation measurements for outdoor urban mobile and backhaul communications in New York City," in *Proc. IEEE Int. Conf. Commun. (ICC)*, Jun. 2014, pp. 4862–4867.

[20] T. S. Rappaport, E. Ben-Dor, J. N. Murdock, and Y. Qiao, "38 GHz and 60 GHz angle-dependent propagation for cellular & peer-to-peer wireless communications," in *Proc. IEEE Int. Conf. Commun. (ICC)*, Jun. 2012, pp. 4568–4573.

[21] P. D. Katev, "Propagation models for WiMAX at 3.5 GHz," in *Proc. ELEKTRO*, May 2012, pp. 61–65.

[22] M. Hata, "Empirical formula for propagation loss in land mobile radio services," *IEEE Trans. Veh. Technol.*, vol. 29, no. 3, pp. 317–325, Aug. 1980.

[23] T. A. Thomas et al., "A prediction study of path loss models from 2–73.5 GHz in an urban-macro environment," in *Proc. IEEE 83rd Veh. Technol. Conf. Spring (VTC-Spring)*, May 2016, pp. 1–5.

[24] T. S. Rappaport, R. W. Heath, Jr., R. C. Daniels, and J. N. Murdock, *Millimeter Wave Wireless Communications*. Englewood Cliffs, NJ, USA: Prentice-Hall, 2015.

[25] S. Sun et al., "Investigation of prediction accuracy, sensitivity, and parameter stability of large-scale propagation path loss models for 5G wireless communications," *IEEE Trans. Veh. Technol.*, vol. 65, no. 5, pp. 2843–2860, May 2016.

[26] S. Sun and T. S. Rappaport, "Multi-beam antenna combining for 28 GHz cellular link improvement in urban environments," in *Proc. IEEE Global Commun. Conf. (GLOBECOM)*, Dec. 2013, pp. 3754–3759.

[27] G. R. MacCartney, Jr., M. K. Samimi, and T. S. Rappaport, "Omnidirectional path loss models in New York City at 28 GHz and 73 GHz," in *Proc. IEEE 25th Int. Symp. Pers. Indoor Mobile Radio Commun. (PIMRC)*, Sep. 2014, pp. 227–231.

[28] M. K. Samimi, T. S. Rappaport, and G. R. MacCartney, Jr., "Probabilistic omnidirectional path loss models for millimeter-wave outdoor communications," *IEEE Wireless Commun. Lett.*, vol. 4, no. 4, pp. 357–360, Aug. 2015.

[29] A. Alwarafy, A. I. Sulyman, A. Alsanie, A. Alshebeili, and H. Behairy, "Receiver spatial diversity propagation path-loss model for an indoor environment at 2.4 GHz," in *Proc. IEEE 6th Int. Conf. Netw. Future (NOF)*, Oct. 2015.

[30] M. K. Samimi and T. S. Rappaport, "Statistical channel model with multi-frequency and arbitrary antenna beamwidth for millimeter-wave outdoor communications," in *Proc. IEEE Globecom Workshops (GC Wkshps)*, Dec. 2015, pp. 1–7.

[31] M. K. Samimi, S. Sun, and T. S. Rappaport, "MIMO channel modeling and capacity analysis for 5G millimeter-wave wireless systems," in *Proc. 10th Eur. Conf. Antennas Propag. (EuCAP)*, Apr. 2016, pp. 1–5.

[32] G. R. Maccartney, Jr., T. S. Rappaport, M. K. Samimi, and S. Sun, "Millimeter-wave omnidirectional path loss data for small cell 5G channel modeling," *IEEE Access*, vol. 3, pp. 1573–1580, 2015.

[33] D. J. Love and R. W. Heath, Jr., "Necessary and sufficient conditions for full diversity order in correlated Rayleigh fading beamforming and combining systems," *IEEE Trans. Wireless Commun.*, vol. 4, no. 1, pp. 20–23, Jan. 2005.

[34] M.-S. Alouini and A. J. Goldsmith, "Capacity of Rayleigh fading channels under different adaptive transmission and diversity-combining techniques," *IEEE Trans. Veh. Technol.*, vol. 48, no. 4, pp. 1165–1181, Jul. 1999.

[35] A. I. Sulyman and M. Kousa, "Bit error rate performance of a generalized diversity selection combining scheme in Nakagami fading channels," in *Proc. IEEE Wireless Commun. Netw. Conf. (WCNC)*, vol. 3. Sep. 2000, pp. 1080–1085.

[36] R. K. Mallik, M. Z. Win, and J. H. Winters, "Performance of dual-diversity predetection EGC in correlated Rayleigh fading with unequal branch SNRs," *IEEE Trans. Commun.*, vol. 50, no. 7, pp. 1041–1044, Jul. 2002.

[37] N. Kong and L. B. Milstein, "Average SNR of a generalized diversity selection combining scheme," *IEEE Commun. Lett.*, vol. 3, no. 3, pp. 57–59, Mar. 1999.

[38] G. R. MacCartney, M. K. Samimi, and T. S. Rappaport, "Exploiting directionality for millimeter-wave wireless system improvement," in *Proc. IEEE Int. Conf. Commun. (ICC)*, London, U.K., 2015, pp. 2416–2422.

[39] T. S. Rappaport et al., "Analysis and simulation of interference to vehicle-equipped digital receivers from cellular mobile terminals operating in adjacent frequencies," *IEEE Trans. Veh. Technol.*, vol. 60, no. 4, pp. 1664–1676, May 2011.

[40] S. Sun et al., "Millimeter wave multi-beam antenna combining for 5G cellular link improvement in New York City," in *Proc. IEEE Int. Conf. Commun (ICC)*, 2014, pp. 5468–5473.

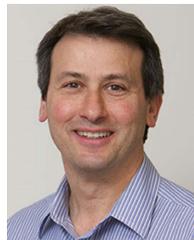

**Theodore S. Rappaport** (S'83–M'84–SM'91–F'98) received the B.S., M.S., and Ph.D. degrees from Purdue University, West Lafayette, IN, USA, in 1982, 1984, and 1987, respectively, all in electrical engineering. He is an Outstanding Electrical and Computer Engineering Alumnus and a Distinguished Engineering Alumnus from his alma mater. He is the David Lee/Ernst Weber Professor of Electrical and Computer Engineering with the New York University Tandon School of Engineering, New York University (NYU), Brooklyn, NY, USA, and is the Founding Director of the NYU WIRELESS Research Center. He also holds professorship positions with the Courant Institute of Mathematical Sciences and the School of Medicine, NYU. He founded major wireless research centers with the Virginia Polytechnic Institute and State University (MPRG), The University of Texas at Austin (WNCG), and NYU (NYU WIRELESS) and founded two wireless technology companies that were sold to publicly traded firms. He is a highly sought-after technical consultant having testified before the U.S. Congress and having served the ITU. He has advised more than 100 students, has more than 100 patents issued and pending, and has authored or co-authored several books, including the best-selling books entitled *Wireless Communications: Principles and Practice—Second Edition* (Prentice Hall, 2002). His latest book entitled *Millimeter Wave Wireless Communications* (Pearson/Prentice Hall, 2015) is the first comprehensive text on the subject.

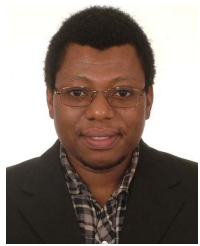

**Ahmed Iyanda Sulyman** (SM'09) received the Ph.D. degree from the Department of Electrical and Computer Engineering, Queen's University, Canada, in 2006. He was a Teaching Fellow with Queen's University from 2004 to 2006 and a Post-Doctoral Fellow with the Royal Military College of Canada from 2007 to 2009. He joined the Department of Electrical Engineering, King Saud University, Saudi Arabia, in 2009, where he is currently an Associate Professor. He has authored over 70 technical papers, six book chapters, and a book entitled *Nonlinear MIMO Communication Channels* (LAP LAMBERT Academic Publishing, Saarbrucken, Deutschland/Germany: CRC Press, 2012). His research interests are broadly in wireless communications and networks, with most recent contributions in the areas of millimeter-wave 5G cellular technologies and the Internet of Things. He has been the Session Chair and Technical Program Committee Member in many top-tier IEEE conferences, including the most-recent IEEE-ICC conference in 2016.

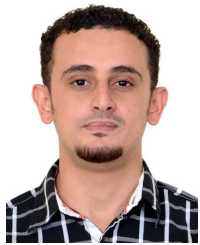

**Abdulmalik Alwarafy** received the B.S. degree in electrical engineering with a minor in communication from IBB University, IBB, Yemen, in 2009, and the M.Sc. degree in electrical engineering with a minor in communication from King Saud University, Riyadh, Saudi Arabia, in 2015, where he is currently pursuing the Ph.D. degree in electrical engineering with the College of Engineering. His research interests include millimeter-wave measurements and propagation channel modeling for the fifth-generation cellular mobile communications.

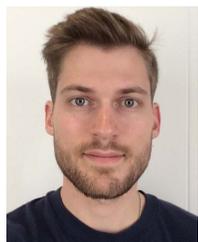

**George R. MacCartney, Jr.** (S'08) received the B.S. and M.S. degrees in electrical engineering from Villanova University, Villanova, PA, USA, in 2010 and 2011, respectively. He is currently pursuing the Ph.D. degree in electrical engineering with the New York University Tandon School of Engineering, New York University, Brooklyn, NY, USA, under the supervision of Prof. Rappaport with the NYU WIRELESS Research Center. He has authored or co-authored over 20 technical papers in the field of millimeter-wave (mmWave) propagation. His research interests include mmWave propagation test and measurement prototyping, and channel modeling and analysis for fifth-generation communications.

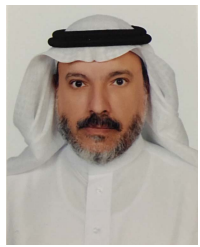

**Abdulhameed Alsanie** received the B.Sc. (Hons.) and M.Sc. degrees from King Saud University in 1983 and 1987, respectively, and the Ph.D. degree from Syracuse University, NY, USA, in 1992, all in electrical engineering. He is an Associate Professor and the Head of Electrical Engineering Department, King Saud University. His current research interests include wireless communications with a focus on multiple input multiple output wireless systems, space time codes, and cooperative wireless systems.